\documentstyle[11pt]{article}
\begin{document}
\begin{center}
\large{
Einstein Field Equation: the Root of All Evil? Quantum Gravity, Solar Application, and Prediction on Gravity Probe B
}
\normalsize\\
Jin He \\
Department of Mathematics, Huazhong University of Science and Technology, \\Wuhan, 430074, P.\,R.\,China\\
E-mail:mathnob@yahoo.com\\
\mbox{   }
\end{center}
\large {{\bf Abstract} } \normalsize
The starting point of quantum mechanics is the  relationship between
energy and momentum: energy is proportional to the squared momentum!
As a result, energy and momentum have not been treated equally. The
wave equation required by quantization is a differential equation.
Quantization means that the energy quantity is replaced by the
derivative to time while the momentum quantity is replaced by the
derivative to space. Both have a common factor which is exactly the
Planck constant. Same to the formula of energy and momentum, the
resulting differential equation does not treat time and space
equally. As a result, the Planck constant is  not canceled out from
the two sides of the differential equation. That is, Planck constant
(the common factor) remains and is the constant which describes the
micro-world.

Dr. He's 
      gravity is the local bending of background space-time which,
 as suggested by Einstein, can be described by
a differential form of second order which treats time and space
equally. Therefore, the Planck constant is completely canceled out
in the resulting differential equation. In other words, the
quantization of gravity simply does not need the Planck constant!
Quantum gravity is even simpler than the familiar quantization of
micro-world: there is no Planck constant!  This is because gravity
obeys Equivalence Principle.
\\
\\
keywords: Relativity -- Gravity Probe B –-  Gravitational Theory -- Quantization : Hydrogen Atom
\\
\\
 \section{ The Most Basic Problem about the Universe and my Answer}

{\sf The basic fact is that there exist structures in the universe.
Stars and planets form the most basic and magnificent systems
(galaxies) in the universe while atoms and molecules form the
vulnerable yet most advanced bodies (human beings). These natural
systems, either vulnerable or magnificent, primitive or advanced,
are in essence composed of smaller parts and present in orderly and
varying existence. The most basic problem about the universe is how
the structures come. Current foundational scientific theories are
incompetent. They can not provide any basic principle to explain
such harmonic structure as human being, nor to resolve the motion of
the most simple physical systems (such as interactional three
bodies). The most simple and primitive system is the one of free
three-bodies which move under pure gravitational force. We call it
self-gravitational three-bodies. However, human beings have not
known how the three bodies move and where the bodies should take
positions to achieve their natural and harmonic configuration.

My answer is that the structure of the universe results from
gravity. The origin of natural structure can not be other forces.
Modern science has fully proved that independent system of
microscopic particles combined by electromagnetic force or nuclear
forces inevitably moves towards chaotic state rather than orderly
one. This is the principle of entropy increase, which is well known
for scientists. Therefore, if there were no gravitational force then
the whole universe would be simply uniform gas without structure.

In the large-scale galaxy system, gravity demonstrates itself as
rational proportion and impulsive disturbing waves. In local
structure of the universe, gravity demonstrates itself as
quantization. Newton and Einstein's gravitational theories are the
demonstration of local rational gravity. There exists local
impulsive gravity too. However, gravity is very very weak to be
experimentally studied. For example, it is $0.0000 \cdots 00001 $
times (where 40 zeros are after the decimal point) weaker than the
electricity! Only the Earth, Moon, Sun and so on present gravity.
There is no slightest gravity between cars or human bodies.
Therefore, human beings suffer insurmountable difficulty to
experimentally study gravity.

Based on Einstein general relativity, gravity has not been
quantized. According to
  Dr. He's 
  theory of gravity (local
bending of flat spacetime), gravity is successfully quantized. This
involves two critical questions. One is why Planck constant exists
and the other is if gravity is completely relative as Einstein
suggested. We will study the questions and give their simple answers
in next Sections. }

 \section{ My Quantum Gravity }

{\sf Einstein's theory does not allow gravity to be quantized. The
people who really understand the principle of quantum mechanics
know: quantization is based on a background causal relation. This
causal relation is exactly the background reference frame. In other
words, the target which is to be quantized must stay in a
background, and be independent of the background. All physical
theories except Einstein general relativity claim that there is a
background inertial reference frame, and as a result, are
successfully quantized. However, Einstein hated inertial reference
frames. According to general relativity, gravity is the bending
background space-time itself.
   Dr. He's 
    theory of gravity is the
local bending of globally flat background space-time. In other
words, far away from the local mass and energy, space-time is
becoming flat. This requires that the large-scale universe be flat.
But Einstein general relativity involves no reference frame not to
mention the flat reference frames, and his gravity is the causal
relation itself. To quantize Einstein gravity is to directly
quantize the causal relation. This, of course, is a failure.

        Dr. He's 
          theory of gravity is also curved space-time. But it is the
local bending of background flat space-time. In other words, away
from the local bending area, space-time is becoming flat and serves
the background causal relation! As a result,
  Dr. He's 
   theory of gravity can
be quantized and the planetary distribution of the solar system (the
Titius-Bode law) can be explained.

The starting point of quantum mechanics is the  relationship between
energy and momentum: energy is proportional to the squared momentum!
As a result, energy and momentum have not been treated equally. The
wave equation required by quantization is a differential equation.
Quantization means that the energy quantity is replaced by the
derivative to time while the momentum quantity is replaced by the
derivative to space. Both have a common factor which is exactly the
Planck constant. Same to the formula of energy and momentum, the
resulting differential equation does not treat time and space
equally. As a result, the Planck constant is  not canceled out from
the two sides of the differential equation. That is, Planck constant
(the common factor) remains and is the constant which describes the
micro-world.

Dr. He's 
      gravity is the local bending of background space-time which,
 as suggested by Einstein, can be described by
a differential form of second order which treats time and space
equally. Therefore, the Planck constant is completely canceled out
in the resulting differential equation. In other words, the
quantization of gravity simply does not need the Planck constant!
Quantum gravity is even simpler than the familiar quantization of
micro-world: there is no Planck constant!  This is because gravity
obeys Equivalence Principle. }

\section{ Newton and Einstein's theory can not explain the distribution of
planets in the solar system }

{\sf As long as a problem involves three or more free bodies, Newton
and Einstein's gravitational theories are powerless. The simplest
example is the distribution of planets in the solar system. Is the
distribution of planets in the solar system orderly? Is the universe
meaningful? Is human life the orderly result?

The solar system is a planar distribution of planets and asteroids.
All planets and asteroids move at almost circular orbits and the
orbits center at the Sun. The Titius--Bode law is a hypothesis that
the planets and asteroids orbit at the exponential series of radii.
The law relates the radius, $a$, of each planet outward from the sun
in the unit such that the Earth's radius is 1, and the formula of
the law is
       $$ a = (n +4) / 10 $$
where $n = 0, 3, 6, 12, 24, 48 ... $, and each value of $n > 3$ is
twice the previous one: $6 = 2 \times  3, 12 = 2 \times  6, 24 = 2
\times 12, 48 = 2 \times  24, $ etc. Here are the distances from the
Sun of all planets calculated with the formula and their real ones: }

\newpage
\begin{table}
\caption{Radii of Planetary Orbits in the Solar System }
\begin{tabular}{lll}
\hline
Planet    & Real radius  &  T-B law \\
   &     &              \\
    Mercury        &        0.39  &                   0.4  (n = 0)\\
    Venus           &       0.72   &                 0.7  (n = 3)\\
    Earth            &      1.00    &                1.0  (n = 6)\\
    Mars              &     1.52     &               1.6  (n = 12)\\
    Asteroid belt      &    2.90       &             2.8  (n = 24)\\
    Jupiter            &    5.20       &             5.2  (n = 48)\\
    Saturn              &   9.54        &            10.0 (n = 96)\\
    Uranus               &  19.18        &           19.6 (n = 192)\\
    Neptune               & 30.06         &          38.8 (n = 384) \\
\hline
\end{tabular}
\end{table}

{\sf From the Table we see that the distribution of planets and
asteroids in the solar system  is meaningful and can be expressed by
simple formula. However, Newton and Einstein's gravitational
theories are no more than the theory of free two-bodies, which can
not explain the orderly co-existence of many bodies. }

\newpage

\begin{figure}[b]
\hspace{4cm}
 \vspace{0cm}
 \includegraphics{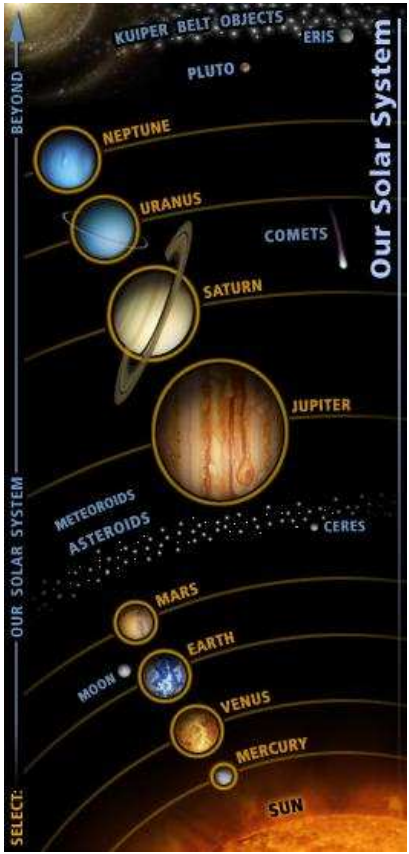}
\caption[]{
 The relative sizes and logarithmic distances of the planets against the Sun in the solar system (image credit
  \cite{he90}).    }
\end{figure}

\newpage
\mbox{  }
\newpage


\section{ Laurent Nottale's Quantum Explanation of Solar System }

{\sf Is the solar system an accidental structure? Mainstream
astrophysicists' answer is yes because Einstein general relativity
can not explain the structure. However,  shortly after the quantum
wave function of Hydrogen atom was proposed, people found out that
the function could be used to explain the structure of the solar
system. Of course, the physical parameters  contained in the
hydrogen atom wave function must be replaced by some parameters
describing the macroscopic world. In particular, the Planck constant
needs to get rid of!

After Planck proposed the quantum concept in 1900, Bohr proposed in
1914 a concept of discrete energy states of hydrogen atom. Quantum
mechanics was established in 1925 to explain the no-orbit motion of
micro-particles with wave functions: the probability wave.

Atomic state is steady state (bound state), and all possible states
correspond to a series of wave functions, denoted by a set of
integer numbers
 $$
  | n, l>
  $$
Where $n$ is the energy quantum number and $l$ is the quantum number
of angular momentum.

French scientist Laurent Nottale considers the nature not only obeys
Einstein's theory of relativity but also the relativity of scales.
Nottale used the hydrogen atom wave function to describe the
structure of the solar system. Of course, the physical parameters
contained in the hydrogen atom wave function must be replaced by
some parameters describing the macroscopic world.  For example, the
Planck constant has been replaced by the scale constant $w_0$ of
scale relativity.

Because the motion of planets in the solar system is approximately
the circular one which centers at the Sun, the angular quantum
number $l$ takes the maximum value (i.e, $l = n-1$). Therefore, the
remaining quantum number describing the solar system is the energy
quantum number $n$.

According to Nottale, \\
the quantum number of Mercury is $n = 3$, \\
the quantum number of Venus is $n = 4$, \\
the quantum number of Earth is $n = 5$, \\
the quantum number of Mars is $n = 6$.

In Fig.\,2, M stands for Mercury, E for Earth, and so on. Hun, C,
Hil, and so on are the asteroids in the asteroid belt. The
horizontal axis in Fig.\,2 measures the quantum number, and the
vertical axis measures the corresponding radii of the planets'
orbits (i.e, the distances from the Sun). Nottale suggests that
there exist two planets of small masses between the Sun and Mercury
corresponding to the quantum number $n = 2$ and $n = 1$
respectively.

Because angular quantum number is $l = n-1$, each wave function has
only one extreme point (maximum point, see Fig.\,3). According to
Nottale, the solar system is a hierarchical structure. The asteroid
belt is between Mars and Jupiter. We note from Fig.\,2 that
Asteroids and the four planets closest to the Sun have very small
masses. According to Nottale's theory of scale relativity, this set
of small planets and asteroids forms the first level of the
hierarchical structure. Nottale calls it the internal level, denoted
by ISS or IS (see Fig.\,2). The scale value of the level is $w_0$.

The solar system has the second level: the external level, denoted
as OSS or OS (see Fig.\,2). The scale value of the level is
$w_0/5$. The scale parameter corresponds to the Planck constant in
microscopic world but varies with the macroscopic levels. The
maximum points of the wave functions in the OSS level correspond to
the positions of the planets (radii): \\
the quantum number of ISS is $n = 1$, \\
the quantum number of  Jupiter is $n = 2$, \\
the quantum number of Saturn is $n = 3$, \\
the quantum number of Uranus is $n = 4$, \\
the quantum number of Neptune is $n = 5$, \\
the quantum number of Pluto is $n = 6$. \\
Note that the mass center of the first level (i.e, the internal
level) is the first ``planet'' of the second level (i.e, the
external level) with the quantum state $n = 1$. This is the
hierarchical structure!

The above is the prediction on planets' distances to the Sun. The
maximum points correspond to the distances. Because the wave
functions of scale relativity are not required to be normalized, the
values of the wave amplitudes have the realistic meaning too. They
correspond to the masses of the planets (see Fig.\,3). Therefore,
Nottale's theory has a full account to the solar structure.

Because the sum of amplitudes of all wave functions in the internal
level is equal to the amplitude of the first wave function in the
external level (see Fig.\,3), the mass distribution of the solar
system is really a hierarchical structure (see Figures \,4 and
5). My quantum gravity has justified Nottale's results
consistently. }

\newpage
\begin{figure}[b]
\hspace{8cm}
 \vspace{0cm}
 \includegraphics{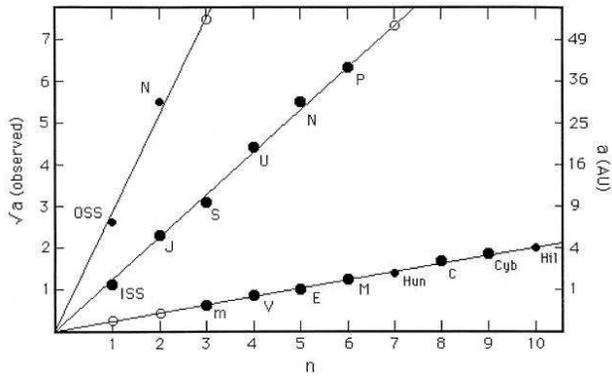}
\caption[]{Comparison of the observed average distances of planets
from the Sun with the theoretical values \cite{he91}. On the inner
system, one has Mercury (M), Venus (V), the Earth (E), Mars (M), and
the main mass peaks of the asteroids belt: Hungarias (Hun), Ceres
(C), Hygeia (Hyg) and Hildas (Hil).    }
\end{figure}

\newpage
\mbox{  }
\newpage

\newpage
\begin{figure}[b]
\hspace{8cm}
 \vspace{0cm}
 \includegraphics{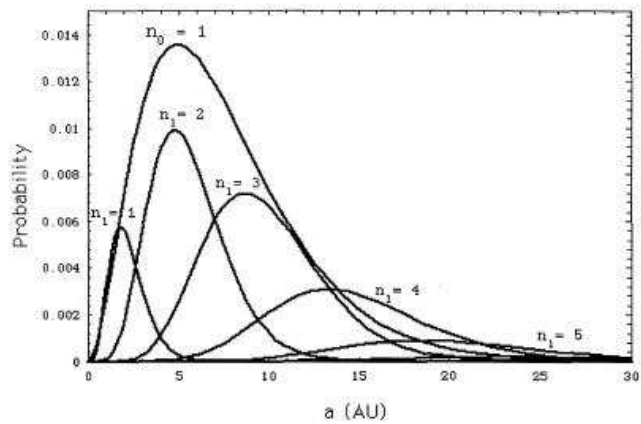}
\caption[]{Schematic representation of the hierarchy process. The
orbital $n_0 = 1 $ is divided into sub-orbitals $n_1 = 1$ (inner
system), $n_1 = 2$ (Jupiter), $n_1 = 3$ (Saturn). The same is true
for the inner system that fragments itself into orbitals $n_2 = 1,2,
3$ (Mercury), 4 (Venus) (\cite{he91}).    }
\end{figure}

\newpage
\mbox{  }
\newpage

\newpage
\begin{figure}[b]
\hspace{8cm}
 \vspace{0cm}
 \includegraphics{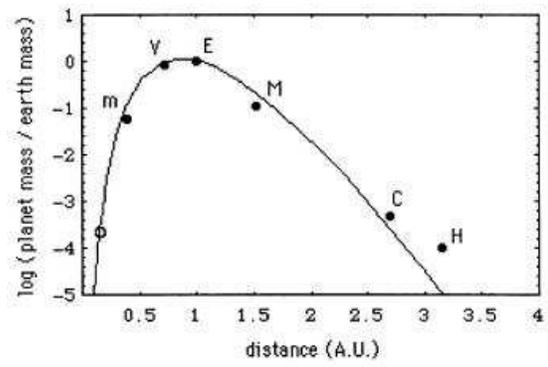}
\caption[]{Comparison of the predicted and observed masses of
planets for the inner solar system. C and H stand for the mass peaks
in the asteroid belt (Ceres and Hygeia). The possible additional
planet (open circle) is expected to have a mass of about
$10^{-4}m_{{\rm earth }}$ (\cite{he91}).    }
\end{figure}

\newpage
\mbox{  }
\newpage

\newpage
\begin{figure}[b]
\hspace{8cm}
 \vspace{0cm}
 \includegraphics{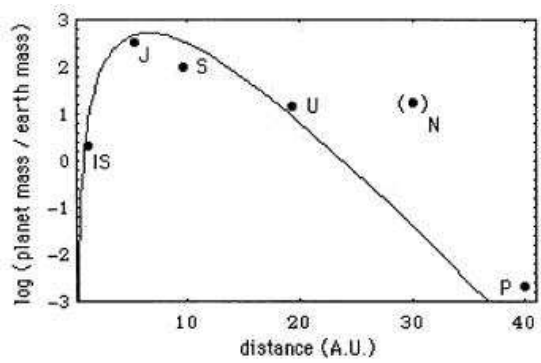}
\caption[]{Comparison of the predicted and observed masses of
planets for the outer solar system. IS stand for the inner system as
a whole. Only Neptune is discrepant, which could be the signature of
another hierarchic structure at larger scale (\cite{he91}).    }
\end{figure}

\newpage
\mbox{  }
\newpage

 \section{  Everything is dependent }

{\sf What is the most important thing in human life? There is no
answer to the question throughout the human history. This is because
humans do not know their own origin. Now the study on the origin of
galaxies tells us the answer. That is, the most important thing in
human life is the gravity. Without the gravity humans would die as
the astronauts would die if they stayed in International Space
Station for over two years.
Therefore, everything is dependent on gravity.

However, the measurement of gravity is dependent on the reference
frame with respect to which the measuring equipment is static. If
our measuring equipment is static with respect to the buildings on
the Earth then the astronauts in International Space Station do
suffer gravity from the Earth.  The result that the astronauts
suffer no gravity holds only with respect to the International Space
Station itself. If we stay in International Space Station and watch
particle movement inside the Station then we can see that all
particles move straightly at constant speeds. That is, all particles
and the astronauts suffer no gravity.

However, if we stay in the building on the ground and watch particle
movement inside the building then we can see that all particles go
parabolic motion if they do not suffer any non-gravitational
interaction. That is, all particles suffer the gravity from the
Earth.

If you stand on the ground, the ground gives supporting force in
balance of the Earth's gravity. Therefore, human cells suffer the
force of pressure from each other from bottom to top. The cells of
astronauts in International Space Station, however, suffer no such
pressure. Therefore, the astronauts suffer a 22\% blood volume
decrease or loss within 2 to 3 days. Muscle atrophy occurs at 5\%
per week. Bone atrophy occurs at 1\% per month. Remember, our bodies
were developed within the earth's gravity field, our muscles are
needed to counter gravity. In space, we no longer need our muscles,
so the brain starts to optimize the body by getting rid of what it
doesn't need.

In summary, to solve any physical problem, we need firstly to
determine the reference frame with respect to which the measuring
equipment is static. Secondly, we need to identify the particles
whose motion is to be measured.

However, Einstein made a mistake in his general theory of
relativity. His theory has no reference frame! What Einstein did is
against the Mach Principle. }

\section{ Mach Principle }

{\sf The above Section shows that everything is dependent on its
environment and human life depends on the gravitational field of
Earth. Mach Principle says the same thing. For example, when you
stand on the ground and relax, your arms fall down naturally.
However, if you rotate your body then your arms are lifted up as the
rotation is faster and faster. Newton thought that the static state
of body corresponded to the absolute space. He provided further
experiment for the argument of the absolute space. That is Newton's
famous water bucket experiment. The experiment works like this:

Assume the bucket which is filled with water is rotated. At first
the water inside it does not move, and the surface of the water is a
plane as if the bucket were not rotated. But at last there appears a
concave on the water surface when the water inside the bucket is
rotated together with the bucket. The experiment indicates that the
water surface remains a smooth plane when the water is immobile no
matter whether the water moves relatively to the bucket or not. But
the water surface remains a concave when the water rotates
regardless of its state relative to the bucket.

In the face of such a conflict, Newton said that, this just proved
the existence of an absolute space. According to the water surface
being flat or concave, we can determine whether the water is
immobile or rotational relative to the absolute space.

Mach is not content with Newton's argument because no evidence can
show that the experiment is just a comparison between two relative
motions rather than the motions in the relative space and the
absolute space.

Many people expressed the same objections before Mach that cast
shadow on Newton's idea of the absolute space. But Mach thinks more
than that. He designed a thought experiment to deepen the
understanding of Newton's bucket, that is, if we fix Newton's bucket
and let the stars and the heaven revolve about it, what can happen
next, can the centrifugal forces appear and the concave of the
surface of the water is formed again?

If the situation remains the same as what it would be when we fix
the remote stars and rotate the bucket, then the logic is that there
must be some influence or forces on the water in the bucket exerted
by the remote stars. On the contrary, if the surface of water
remains horizontal when the stars are rotating, that means Newton's
proof of the existence of the absolute space is correct. This
experiment can only be done in mind, but the answer to this question
is rather baffling.

Mach principle is that the matter of the whole universe can affect
local dynamical systems. That is, the matter of the whole universe
sets up the local absolute reference frames.

However, Einstein general theory of relativity is against Mach
Principle. How does relativity coexist with the local absolute
reference frame. Relativity becomes a religion governing the whole
universe. Any attempt to exploring the origin of the human race and
the universe is blocked by the religion.

However, the existence of the local inertial reference frame is a
fact. My results in following show that Einstein's general theory of
relativity can coexist with the absolute reference frame. Based on
the absolute reference frame, gravity can be quantized as indicated
in the following Sections. }

\section{ Quantum Gravity of the Sun }

{\sf Mach principle tells us that, if our hands are unconsciously
lifted, it is that our body is rotating about the space of material
background. Mach principle means that everything is dependent on its
material background. This is the common sense which common people
may neglect. However, the uncommon giant Einstein neglected it too!

Human life depends on Earth and its environment. Human body is the
product of the Earth and its cosmic environment, and can be
explored. However, the general theory of relativity blocks any
attempt to the exploration.

Now I introduce my quantum gravity (i.e., the quantum gravitational
field of the Sun) which is based on the assumption of local bending
of flat spacetime. The local bending of flat spacetime is the
modification of general relativity according to Mach Principle.

The general theory of relativity is just a sentence: gravity is the
curved space-time. My theory of gravity is the curved space-time
too, but I recognize the Mach principle. I admit that everything is
dependent on its material background. That is, I admit the existence
of background flat space-time. The Sun with its huge mass and energy
warps its flat background. Therefore, my gravity is the local
bending of flat background space-time. The space and time on the
flat background play the absolute role. But general relativity has
no absolute space!

Anyone who has learned a little physics knows that the description
of any physical phenomenon has to involve the coordinates of
space-time. Gravity is the space-time itself (nevertheless, it is a
curved space-time). Therefore, the description of gravity is even
more inseparable from the concept of coordinates.

It is the mathematician Riemann who first put forward the concept of
multi-dimensional bending space. It is the same guy who proved his
most important theorem: in and only in the flat space does there
exist one coordinate system whose coordinates themselves are the
global properties of the space: the distances in the space.

Friends, you are now aware of the disadvantages of bending
spacetime: you can not find a coordinate system in any bending
space-time which describes everywhere the global properties:
distances or time or angles. The most familiar Cartesian coordinates
(distances and time), $x, y, z, t,$ no longer exist in bending
spacetime. The coordinates are nothing but mathematical symbols. To
calculate distances or time in bending space-time, you have to
perform the integration of metrical tensor. This mathematical
technique, however, is something of nightmare for those professors
of theoretical physics who have little mathematical discipline. Even
Einstein did not have such background. It is the mathematician
Marcel Grossmann who helped him set up the mathematical basis of the
general theory of relativity.

Of course, if confined to the curved space-time itself, you can not
find the coordinates which have the direct meaning of distance or
time. However, I respect Mach Principle and I have the absolute
background which provides the absolute Cartesian coordinates of the
absolute distance or time.

I use the external environment of the Sun, i.e., the flat spacetime
to describe Sun's gravity. The Cartesian coordinates of the external
environment are $x, y, z, t$. Therefore, my theory of gravity,
whether it is the Sun's gravitational field or its quantization, is
based on the Cartesian coordinates $x, y, z, t$. This is fully
studied in the following Section. Here is its summary.

Firstly let us see the Lagrange and Hamilton description of the
gravity of a mass $M$. The formula (13) (see the following Section) is the
metric tensor of Einstein curved space-time
$$
\begin{array}{l}
-\frac {1}{2}\grave {s} ^2 = L(x^i, x^0, \grave x ^i, \grave x ^0) \\
=\frac{1}{2} g_{\alpha \beta }\grave x ^\alpha  \grave x ^\beta .
\end{array}
$$
However, the coordinates $x, y, z, t$ is the absolute coordinates of
the flat background which have the direct meaning of distance and
time! My friends, have you noticed the uppercase letter $L$ and $H$
in the Section? They are the Lagrangian and Hamiltonian of the
gravitational field respectively.

There is no reference frame for Einstein's general theory of
relativity not to mention the inertial reference frame.  Einstein
hates inertial reference frames. This is because the inertial
reference frame is an absolute concept which is incompatible with
relativity! However, Einstein's fans including professors have no
ability to understand the differential geometry and have often
written the metric tensor as Lagrangian just as what I do. You know,
if there were no reference frame, there would be no concept of
energy, no Lagrangian and Hamiltonian! This is the clumsy performance
of Einstein's followers.

However, I use the coordinates of the flat background space-time to
describe the curved space-time. Therefore, the use of Lagrangian and
Hamiltonian is legal! If the gravitational field is weak, the
Lagrangian and Hamiltonian do return to the classical Lagrangian and
energy!

The formula (16) is the well-known vacuum solution of Einstein field
equation, i.e., the \\
Schwarzschild metric
$$
\begin{array}{l}
-\frac {1}{2}\grave {s} ^2 =-\frac{1}{2} (1-2r_g/r ) c^2\grave
t^2 \\
+ \frac{1}{2}\grave r^2/(1-2r_g/r ) +\frac{1}{2}r^2( \grave \theta
^2 +\sin ^2\theta \grave \phi ^2  ).
 \end{array}
$$
Its only parameter is $r_g$ in (15)
$$
r_g(=GM/c^2)
$$
This parameter is called the Schwarzschild radius. For the Earth,
this radius is equal to about 0.001 meters, or the magnitude of
millimeters. Because the Earth's physical radius is 6.378 million
meters and humans live on the Earth radius, the Schwarzschild radius
divided by the Earth's radius is about  0.000,000,000,01. Therefore,
the above formula is approximately the metric of flat space-time,
i.e., the un-definite Euclidean metric. In other words, the bending
of the space-time where humans live is 0.000,000,000,01!

From now on, we divide the Schwarzschild radius of the involved mass
(e.g., the Earth or the Sun) by the physical radius of the mass, and
the result is called the bending parameter of curved spacetime.
Because the bending parameters of Earth and Sun are both very small,
current astronomical observations can confirm Einstein theory in its
first-order approximation with respect to the parameter. Therefore,
any declaration that Einstein general theory of relativity is fully
proved is a lie! If you provide a theory and its first-order
approximation shares the same result as general relativity then you
are as great as Einstein!

However, the time delay effect of gravity, i.e., the Shapiro time
delay effect, is an exception, which involves the second-order
approximation. The effect is that a reflected radar signal by a
massive object (e.g., the planet Mercury) takes more travel time
than it would if the object did not exist. However, the observation
of the effect is very complicated and needs to know the orbital
motion under gravity. It does not matter. Shapiro assumed that
general relativity be correct and used general relativity to
calculate the orbit. He claimed that the observation data is
consistent with the prediction of general relativity! You can NOT
assume something be correct and declare it be correct.

The formula 14 is a metric whose first-order prediction is the
same as Schwarzschild metric but the prediction of time delay effect
with respect to Mercury is 260 microseconds while the prediction of
Schwarzschild metric is 240 microseconds. The metric is called the
quantum-solvable metric.

Why should I find a quantum-solvable metric? I tell you: curved
space-time is not a simple effect. It is a highly non-linear effect.
Non-linear differential equations generally have no analytic
solutions! Einstein field equation is highly nonlinear. Currently
only three or four solutions are found. I want to quantize gravity
and I need analytic solution. That is why I provided the
quantum-solvable metric. The following quantization of solar gravity
is based on the metric.

Because the coordinates in the metric are the Cartesian coordinates
on flat background, we can follow the common way of classical
mechanics to seek the corresponding Hamiltonian. Then we follow the
classical method of quantum mechanics to seek the canonical
coordinates and the corresponding canonical momentum. Replacing the
canonical coordinate and momentum with the corresponding operators,
we get the wave equation of quantum gravity.

The formulas (22) through (26) are the general quantization
description for the quantum-solvable metric. The symbol with the
``goat's horn''
$$
\breve h
$$
is the proportion constant corresponding to the
Planck constant. However, the space-time is covariant and the
quantities of time and space have been treated equally. The
proportion constant cancels out completely from the wave equation of
quantum gravity. This means that quantum gravity has no need of the
Planck's constant! Quantum gravity is simpler than the quantum
mechanics of microscopic world. We have less constants!

However, this constant may be the scale parameter $w_0$ of Nottale's
scale relativity.

The formulas (36) through (42) are the result of the general formula
applied  to the gravity of Sun.  The resulting quantum wave function
turns out to be the wave function of hydrogen atom with the Planck
constant being replaced by the scale parameter $w_0$. This confirms
the prediction of more than 80 years ago that solar planetary
distribution obeys the quantum states of hydrogen atom. This is
absolutely not a coincidence. Human body is most likely the result
of quantum gravity! }

 \section{ Quantum Gravity of the Sun }

\subsection{ Covariant Description of Vanishing Gravity}

{\sf {\bf (i) Lagrangian.}

This Section deals with gravitational interaction only, no other
interaction being involved. Newton's first law of motion that a
particle in a vanishing gravitational field must move in straight
direction with constant (or zero) velocity with respect to any
inertial frame $txyz $, can be described by the language of
relativity by introducing Minkowski metric $\eta _{\alpha \beta }$
and proper distance $s$ to the frames,
\begin{equation}
\begin{array}{ll}
-\frac {1}{2}\grave s ^2 &= L(x^i, x^0, \grave x ^i, \grave x ^0) \\
&=\frac{1}{2}( -c^2\grave t^2 + \grave x ^2 +\grave y ^2 +\grave z
^2 ) \\
&=\frac{1}{2} \eta _{\alpha \beta }\grave x ^\alpha  \grave x ^\beta
\end{array}
\end{equation}
where
\begin{equation}
\grave x ^i =\frac{dx^i }{dp}, \mbox{ }{\rm etc.,}
\end{equation}
$p$ is the curve parameter in the 4-dimentional flat spacetime which
can be chosen $\propto s$ (an invariant parameter, or $\propto \bar
s$, another invariant parameter in the following) , $x ^0=ct , x
^1=x ,x ^2 =y , x ^3 =z $, $c$ is light speed, and $ \eta _{00}=-1,
 \eta _{11}= \eta _{22} = \eta _{33} =1, \eta _{\alpha \beta } = 0 (\alpha \not= \beta) $.
Note that, from now on, all letters with grave accent (e.\,g.,
$\grave x$) denotes the derivative of the quantity (e.\,g. $ x$)
with the curve parameter $p$. This is to be distinguished from the
dot accent (e.\,g., $\dot x$) which is the derivative of the
quantity (e.\,g. $x$) with time $t$ (a common notation). The array
$\eta _{\alpha \beta } $  is the Minkowski metric which is the basis
of special relativity. I call the distance $s$ along the curves of
spacetime by real distance. The distance is generally called proper
distance which can be negative because the matrix $\eta _{\alpha
\beta }$ is indefinite.  The indefinite quadratic form (1) is the
generalization of Pythagoras theorem to Minkowski spactime.

It is straightforward to show that the first Newton law of motion
(vanishing gravity) is the result of variation principle applied to
the Lagrangian $L$ (the formula (1)). That is, I need to prove
that the motion of straight direction with constant velocity from
$p_A$ to $p_B$ is such that the line integral,
\begin{equation}
I=\int ^{p_B}_{p_A} L dp,
\end{equation}
is an extremum for the path of motion. The resulting Lagrange's
equation is
\begin{equation}
\frac {d}{dp} \left (\frac {\partial L}{\partial \grave x^\alpha
}\right )- \frac {\partial L}{\partial x^\alpha }=\pm\frac
{d^2x^\alpha }{dp^2}=0 .
\end{equation}
This is exactly the motion of straight direction with constant
velocity,
\begin{equation}
\begin{array}{l}
\frac {dx^0}{dp} ={\rm constant },  \;\;\; \frac {dx^i}{dp} ={\rm
constant }, \\
 i=1,2,3.
\end{array}
\end{equation}
We can choose $dt/dp=1$. Then
\begin{equation}
\frac {dx^i}{dt} ={\rm constant }=v^i,\;i=1,2,3.
\end{equation}
I have recovered the first Newton's law by introducing the
Lagrangian and Minkowski metric (1). In fact, choosing a new
Lagrangian which is any monotonous function of  the original
Lagrangian results the same Lagrange's equation. For example, choose
the Lagrangian
\begin{equation}
L_1(x^i, x^0, \grave x ^i, \grave x ^0)=\sqrt{-2L}=\grave s.
\end{equation}
The resulting Lagrange`s equation is the same formula (5). The
resulting line integral,
\begin{equation}
\int ^{p_B}_{p_A} L_1 dp =s,
\end{equation}
is the proper distance along the line. Therefore, the Lagrange's
equation describes the path of motion which has shortest proper
distance between the two spacetime points corresponding to
parameters $p_A$ and $p_B$ respectively.

{\bf (ii) Hamiltonian.} To quantize gravity according to common
procedure, however, we need a Lerendre transformation to transform
the Lagrangian to the Hamiltonian which represents classical total
energy. In the present section we deal with the motion of free
particle (vanishing gravity). Now we derive the Hamiltonian of unit
mass based on the Lagrangian (1). The canonical momentums to
$x^\alpha , \alpha =0,1,2,3$ are the following,
\begin{equation}
\begin{array}{l}
P_0 =\frac { \partial }{\partial \grave x^0} L=-\frac { cd t }{dp }  \\
P_i =\frac { \partial }{\partial \grave x^i} L=\frac { dx^i }{dp },
\mbox{ }i=1,2,3
\end{array}
\end{equation}
Therefore, the Hamiltonian of the free particle is
\begin{equation}
\begin{array}{ll}
H &=\grave x^0 P_0 +\grave x^i P_i –- L   \\
&= \frac{1}{2} \sum ^3_{i=1}\left (\frac {dx^i  }{dp}\right )^2-\frac{1}{2} c^2\left (\frac {dt  }{dp}\right )^2  \\
&= \frac{1}{2}\sum ^3_{i=1} P^2_i-\frac{1}{2} P^2_0.
\end{array}
\end{equation}
If we choose
\begin{equation}
\frac{dt}{dp}=1
\end{equation}
then $P^2_0=c^2$ and finally the Hamiltonian (total energy) is
 \begin{equation}
H = \frac{1}{2} \sum ^3_{i=1}\left (\frac {dx^i  }{dt}\right
)^2-c^2.
\end{equation}
We see that the spatial part of the Hamiltonian corresponds to
kinetic energy while the temporal part corresponds to potential
energy. Both energies are constants. The potential energy is $-c^2$
which is chosen to be zero in non-covariant theory. }

\subsection{ Einstein's Metric Form Considered to be the Lagrangian on Flat \\ Spacetime and a Solvable Lagrangian }

{\sf {\bf (i) Geometrization of Gravity (GR).} It is more important
to consider test particle's motion in an inertial frame in which the
particle does experience gravitational force. In the frame, the
particle no longer moves in straight direction with a constant (or
zero) velocity. The motion is described in good approximation by the
Newton's universal law of gravitation which is, however, a
non-relativistic theory and needs to be generalized to give account
for the solar observations which deviate from Newton laws'
calculation. Einstein's general relativity (GR) is the most
important try toward the generalization. The basic assumption of GR
is that gravity is spacetime itself but the spacetime is curved.
Therefore, the gravity of GR is the simple replacement of the above
matrix $\eta _{\alpha \beta }$ by a tensor field $g _{\alpha \beta
}$ whose components are position functions on spacetime, instead of
the constants $\pm 1$,
\begin{equation}
\begin{array}{ll}
-\frac {1}{2}\grave {s} ^2 &= L(x^i, x^0, \grave x ^i, \grave x ^0) \\
&=\frac{1}{2} g_{\alpha \beta }\grave x ^\alpha  \grave x ^\beta .
\end{array}
\end{equation}
Similar to the description in the case of vanishing gravity, the
motion of the particle from $p_A$ to $p_B$ in the gravitational
field is such that the line integral (3) is an extremum for the
path of motion. Therefore, the equation of motion is the
corresponding Lagrange's equation which is exactly the known
geodesic equation given in the general theory of relativity (GR).
However, when we call the Lagrange's equation by geodesic equation,
we mean that the spacetime is curved and the real distance on the
curved spacetime is $\sqrt {-2L} = s$. This kind of explanation of
the Lagrangian is called Einstein's geometrization of gravity and
the coordinates $x, y, z, t$ has no direct meaning of spatial
distance or time interval. The tensor field $g_{\alpha \beta }$ in
(13) is called the metric of the curved spacetime.

My theory of gravity is the curved space-time too, but I recognize
the Mach principle. I admit that everything is dependent on its
material background. That is, I admit the existence of background
flat space-time. The Sun with its huge mass and energy warps its
flat background. Therefore, my gravity is the local bending of flat
background space-time. The space and time on the flat background
play the absolute role. But general relativity has no absolute
space!

It is the mathematician Riemann who first put forward the concept of
multi-dimensional bending space. It is the same guy who proved his
most important theorem: in and only in the flat space does there
exist one coordinate system whose coordinates themselves are the
global properties of the space: the distances in the space. Friends,
you are now aware of the disadvantages of bending spacetime: you can
not find a coordinate system in any bending space-time which
describes everywhere the global properties: distances or time or
angles. The most familiar Cartesian coordinates (distances and
time), $x, y, z, t,$ no longer exist in bending spacetime. The
coordinates are nothing but mathematical symbols. To calculate
distances or time in bending space-time, you have to perform the
integration of metrical tensor. This mathematical technique,
however, is something of nightmare for those professors of
theoretical physics who have little mathematical discipline. Even
Einstein did not have such background. It is the mathematician
Marcel Grossmann that helped him set up the mathematical basis of
the general theory of relativity.

Of course, if confined to the curved space-time itself, you can not
find the coordinates which have the direct meaning of distance or
time. However, I respect Mach Principle and I have the absolute
background which provides the absolute Cartesian coordinates of the
absolute distance or time.

I use the external environment of the Sun, i.e., the flat spacetime
to describe Sun's gravity. The Cartesian coordinates of the external
environment are $x, y, z, t$. Therefore, my theory of gravity,
whether it is the Sun's gravitational field or its quantization, is
based on the Cartesian coordinates $x, y, z, t$. Therefore, all
coordinates in the above or in the following formulas have the
absolute meaning: they are the Cartesian or polar coordinates of the
background spacetime! Therefore, all the formulas have the absolute
meaning which is against the relativity creed!

Therefore, the coordinates $x, y, z, t$ in the formula (13) is the
absolute coordinates of the flat background which have the direct
meaning of distance and time! There is no reference frame for
Einstein's general theory of relativity not to mention the inertial
reference frame.  Einstein hates inertial reference frames. This is
because the inertial reference frame is an absolute concept which is
incompatible with relativity! However, Einstein's fans including
professors have no ability to understand the differential geometry
and have often written the metric tensor as Lagrangian just as what
I do. You know, if there were no reference frame, there would be no
concept of energy, Lagrangian and Hamiltonian! This is the clumsy
performance of Einstein's followers.

However, I use the coordinates of the flat background space-time to
describe the curved space-time. Therefore, the use of Lagrangian and
Hamiltonian is legal! If the gravitational field is weak, the
Lagrangian and Hamiltonian do return to the classical Lagrangian and
energy!

{\bf (ii) A solvable Lagrangian.} From now on, we divide the
Schwarzschild radius of the involved mass
(e.g., the Earth or the Sun) by the physical radius of the mass, and
the result is called the bending parameter of curved spacetime.
Because the bending parameters of Earth and Sun are both very small,
current astronomical observations can confirm Einstein theory in its
first-order approximation with respect to the parameter. Therefore,
any declaration that Einstein general theory of relativity is fully
proved is a lie! If you provide a theory and its first-order
approximation shares the same result as general relativity then you
are as great as Einstein!

Now I give you the quantum-solvable metric:
\begin{equation}
\begin{array}{ll}
-\frac {1}{2}\grave {s} ^2 &= L(x^i, x^0, \grave x ^i, \grave x ^0) \\
&=-\frac{1}{2}(\grave x ^0)^2/D(r )+ \frac{1}{2}D(r )\sum ^{3}_{i=1}
(\grave x ^i)^2
 \\
&=- \frac{1}{2} c^2\grave t^2/D(r ) \\
& + \frac{1}{2}D(r )(\grave r^2 +r^2( \grave \theta ^2 +\sin
^2\theta \grave \phi ^2  ))
\end{array}
\end{equation}
where
\begin{equation}
D(r )=1+2r_g/r,
\end{equation}
$r_g(=GM/c^2)$ is the Schwarzschild radius, $M$ is the central point
mass, and $G$ is the gravitational constant. The Lagrandian is
called solvable Lagrangian because its quantization admits exact
solution. I tell you: curved space-time is not a simple effect. It
is a highly non-linear effect. Non-linear differential equations
generally have no analytic solutions! Einstein field equation is
highly nonlinear. Currently only three or four solutions are found.
I want to quantize gravity and I need analytic solution. That is why
I provided the quantum-solvable metric. The following quantization
of solar gravity is based on the metric.

The Lagrangian is a little different from the Schwarzschild metric
form:
\begin{equation}
\begin{array}{l}
-\frac {1}{2}\grave {s} ^2 =-\frac{1}{2} (1-2r_g/r ) c^2\grave
t^2 \\
+ \frac{1}{2}\grave r^2/(1-2r_g/r ) +\frac{1}{2}r^2( \grave \theta
^2 +\sin ^2\theta \grave \phi ^2  ).
 \end{array}
\end{equation}
For solar system, $2r_g/r \ll 1$ and only the first-order
approximations (in $2r_g/r$) of the metric coefficients are
testified. Therefore, we are interested in the approximations. These
approximations of the Schwarzschild coefficients are $-1\times
2r_g/r$ to the term $ - \frac{1}{2}c^2\grave t^2$, $1\times 2r_g/r$
to the term $ \frac{1}{2}\grave r^2$, and $0\times 2r_g/r$ to the
term $\frac{1}{2} r^2( \grave \theta ^2 +\sin ^2\theta \grave \phi
^2  )$. Simply say, the first-order approximations are (-1, +1, 0).
The same approximations for the solvable Lagrangian coefficients are
$(-1, +1, j)$ where $j=+1$. Therefore, only the approximation to the
third term is different between the Schwarzschild form and the
solvable Lagrangian form ($j=0$ for the Schwarzschild form).

{\bf (iii) Integration of the Lagrange's equations of the solvable
Lagrangian.} Due to the spatially isotopic metric, the test
particle's spatial motions are planar. Therefore, we can choose
$\theta =\pi/2 $ in (14). Particle's motion follows the
corresponding Lagrange's equation. The integration of the temporal
Lagrange's equation is
\begin{equation}
\frac {dt }{dp}=aD( r)
\end{equation}
where $a$ is an arbitrary constant. From now on we choose $a=1$,
\begin{equation}
\frac {dt }{dp}=D( r).
\end{equation}
The integration of the Lagrange's equation which corresponds to
$\phi $ is
\begin{equation}
D( r)r^2\frac {d\phi }{dp}=J.
\end{equation}
Furthermore, we have the integration of the Lagrange's equation
which corresponds to radius $r$, with the help of the other
solutions,
\begin{equation}
D( r)\left (\frac {dr }{dp}\right )^2+\frac {J^2 }{r^2D( r)}-D(r
)c^2=E\:({\rm constant}).
\end{equation}

{\bf (iv) The first-order solar predictions of the solvable
Lagrangian. } While geometrization requires spacetime curvature to
determine gravitational metric (i.e., the Einstein field equation),
the Lagrangian on flat-spacetime has no such constraint. Without the
constraint, we are free to see which kinds of Lagrangians give the
similar predictions as Schwarzschild metric. Because all solar tests
of the Schwarzschild metric are made on its first order
approximation in $2r_g/r$, we consider the above-said first-order
approximations $(f, g, j)$. Dr. He \cite{jin1} shows that any
diagonal effective metric whose first-order approximation is $(-1,
+1, j)$, where $j$ is any real numbers, has the same predictions as
Schwarschild metric, on the deflection of light by the sun and the
precession of perihelia.

However, the prediction on the excess radar echo delay depends on
the choice of $j$. The prediction on the maximum excess delay of the
round-trip to Mercury when it is at superior conjunction with
respect to Sun is
\begin{equation}
(\Delta t)_{{\rm max }}\simeq 19.7 (1 +j+11.2) \; \mu \,{\rm sec }.
\end{equation}
Our solvable Lagrangian corresponds to $j=1$ and its prediction is
$(\Delta t)_{{\rm max }}\simeq 260 \mu \,{\rm sec } $ while the
corresponding result of GR $(j=0)$ is 240 $\mu$\,sec. The difference
is less than 8 percents. However, there is difficulty in the test.
We can transmit radar signals to Mercury at its series of orbital
positions around the event of superior conjunction. The time for
single round-trip is many minutes and an accuracy of the order of
0.1 $\mu $\,sec can be achieved \cite{jin2}. In order to compute an
excess time delay, we have to know the time $t_0$ that the radar
signal would have taken in the absence of the sun's gravitation to
that accuracy. This accuracy of time corresponds to an accuracy of
15 meters in distance. This presents the fundamental difficulty in
the above test. In order to have a theoretical value of $t_0$,
Shapiro's group proposed to use GR itself to calculate the orbits of
Mercury as well as the earth [19-21]. The data of time for the above
series of real round-trips minus the corresponding theoretical
values of $t_0$ presents a pattern of excess time delay against
observational date and was fitted to the theoretical calculations
with a fitting parameter $\gamma $. The group and the following
researchers ``declared'' that, among other similar geometrical
theories of gravity represented by $\gamma $, GR fit the pattern
best. }

\subsection{ Hamiltonian of the Solvable Lagrangian and its Quantization}

{\sf To quantize gravity according to common procedure, however, we
need a Lerendre transformation to transform the solvable Lagrangian
to the Hamiltonian which represents the total energy per unit mass.
The canonical momentums to $x^\alpha , \alpha =0,1,2,3$ are the
following,
\begin{equation}
\begin{array}{l}
P_0 =\frac { \partial }{\partial \grave x^0} L=-\frac { cd t }{dp }/D(r ) \\
P_i =\frac { \partial }{\partial \grave x^i} L=D(r )\frac { dx^i
}{dp }, \mbox{ } i=1,2,3
\end{array}
\end{equation}
Therefore, the Hamiltonian of the system is
\begin{equation}
\begin{array}{l}
H =\grave x^0 P_0 +\grave x^i P_i–- L   \\
=-\frac{1}{2} c^2\left (\frac {dt  }{dp}\right )^2/D(r ) +\frac{1}{2} \sum ^3_{i=1}D(r )\left (\frac {dx^i  }{dp}\right )^2  \\
=-\frac{1}{2}D(r ) P^2_0 + \frac{1}{2}\sum ^3_{i=1} P^2_i/D(r ) .
\end{array}
\end{equation}
We see that the Hamiltonian $H$ equals to the solvable Lagrangian
$L$, $H=L$. We also see that the temporal part of the Hamiltonian is
potential energy while the spatial part is kinetic energy.

Application of the common procedure,
\begin{equation}
\begin{array}{l}
H \rightarrow i\breve h\frac{ \partial }{\partial p}   \\
P^0 \rightarrow i\breve h\frac{\partial }{\partial \tilde t}  \\
P^i \rightarrow –- i\breve h\nabla ,
\end{array}
\end{equation}
produces the quantization of the system, where
\begin{equation}
\tilde t=ct=x^0 .
\end{equation}
The resulting wave equation is
\begin{equation}
\begin{array}{l}
i\breve h \frac{\partial }{\partial p}\Psi (x,y,z, \tilde t,p) =\\
\frac{\breve h^2 D(r )}{2} \frac{\partial ^2}{\partial \tilde
t^2}\Psi -\frac{\breve h^2  }{2D(r )}\nabla ^2 \Psi .
\end{array}
\end{equation}
We consider only the eigenstates of energy,
\begin{equation}
\Psi (x,y,z, \tilde t,p) =\Phi (x,y,z, \tilde t)e^{-iEp}.
\end{equation}
The eigenstates $\Phi (x,y,z, \tilde t)$ satisfy the equation
\begin{equation}
\begin{array}{l}
\frac{2E}{-\breve h}\Phi (x,y,z,\tilde t) =\\
- D(r ) \frac{\partial
^2}{\partial \tilde t^2}\Phi +\frac{1}{D(r )}\nabla ^2 \Phi .
\end{array}
\end{equation}
I absorb the quantization number $\breve h$ into the energy constant
$E$. We can not make experiments to measure the energy of solar
system which depends on the specific theory used. The observable
quantities are the planetary distances and orbital motions. We will
not distinguish energy distributions with constant factors and
constant terms. We denote $2E/\breve h$ by $\tilde E$. Our wave
equation is
\begin{equation}
\begin{array}{l}
-\tilde E\Phi (r,\theta ,\phi, \tilde t) =- D(r )\frac{\partial
^2}{\partial \tilde t^2}\Phi \\
+\frac{1}{D(r )}\left ( \frac{1
}{r^2} \frac{\partial }{\partial r}r^2\frac{\partial }{\partial
r}-\frac{l(l+1)}{r^2} \right ) \Phi .
\end{array}
\end{equation}
We consider only the eigenstates of temporal operator,
\begin{equation}
\Phi (r,\theta ,\phi , \tilde t) =\Phi (r,\theta ,\phi )e^{-i\omega
\tilde t}.
\end{equation}
The eigenstates $\Phi (r,\theta ,\phi )$ satisfy the equation
\begin{equation}
\begin{array}{l}
-\tilde E\Phi (r,\theta ,\phi ) = \omega ^2D(r )\Phi \\
+\frac{1}{D(r )}\left ( \frac{1 }{r^2} \frac{\partial }{\partial
r}r^2\frac{\partial }{\partial r}-\frac{l(l+1)}{r^2}\right ) \Phi .
\end{array}
\end{equation}
The solution of the equation is
\begin{equation}
\begin{array}{l}
\Phi (r, \theta ,\phi) =\chi _{nl}Y_{lm}(\theta ,\phi )/r,   \\
\chi _{nl}( r) = {\rm WhittackerM} (
 \frac {r_g(2\omega ^2+\tilde E)}{\sqrt{-\tilde E-\omega ^2}}, \\
\frac{1}{2}\sqrt{(2l+1)^2-16\omega ^2r_g^2}, 2\sqrt{-\tilde E-\omega
^2 }r )
\end{array}
\end{equation}
The Whittacker function $ {\rm WhittackerM}(\kappa, \mu, z)$ is
connected to Kummer's function, i.\,e. confluent hypergeometric
function $\mbox{ }_1F_1(\alpha , \gamma , z)$, by the following
formula
\begin{equation}
\begin{array}{l}
{\rm WhittackerM}(\kappa, \mu, z)= \\
e^{-z/2}z^{\mu +1/2} \mbox{
}_1F_1(\mu+1/2-\kappa, 1+2\mu, z).
\end{array}
\end{equation}
For the wave function to be finite at infinity $r=+\infty $, the
confluent function must be a polynomial, that is,
\begin{equation}
\begin{array}{l}
\frac{1}{2}\sqrt{(2l+1)^2-16\omega ^2r_g^2}+\frac {1}{2}\\
-
 \frac {r_g(2\omega ^2+\tilde E)}{\sqrt{-\tilde E-\omega ^2}}=-n_r
 \end{array}
\end{equation}
where $n_r$ is any non-negative integer. This is the quantization
condition for our system of macrophysics. The solution of the
condition is
\begin{equation}
\begin{array}{l}
\sqrt{-\tilde E-\omega ^2} = \\
\frac{1}{2 r_g} (-\left (n_r+\frac
{1}{2}+\frac{1}{2}
 \sqrt{(2l+1)^2-16\omega ^2r_g^2}\right )   \\
+ \sqrt{4\omega ^2r_g^2+\left (n_r+\frac {1}{2}+\frac{1}{2}
 \sqrt{(2l+1)^2-16\omega ^2r_g^2}\right )^2} ).
\end{array}
\end{equation}

Note that our wave function depends on the spatial quantum numbers
$n_r,l$, temporal quantum number $\omega $, and the gravitational
strength $r_g$. The quantization number $\breve h$ is not involved,
because our Lagrangian is homogeneous of the generalized velocity
components, $ dx^\alpha /dp$ (see (13) and (29)). }

\subsection{ Application to Solar System }

{\sf {\bf (i) Energy levels in the first order approximation. } From
equation (32), the peaks of probability density start at the
``Bohr radius''
\begin{equation}
a_0\propto 1/(2\sqrt{-\tilde E-\omega ^2 }).
\end{equation}
The closer planets to the Sun, e.\,g. Mercury, have approximately
such a radius. Because $2l+1\geq 1$ and $r_g \approx 1.5 \times
10^3$\,m, the quantity $\omega ^2r_g^2 $ must be very small, $\omega
^2r_g^2  \ll 1$,  so that the reciprocal of (34) approaches the
required radius. This requirement leads to
\begin{equation}
\sqrt{-\tilde E-\omega ^2} =\frac{\omega ^2r_g }{4(n_r+l+1)}
\end{equation}
in the first order approximation of $\omega ^2r_g^2 $. This is the
energy level formula for solar system. As usual, we use the quantum
number $n=n_r+l+1$,
\begin{equation}
\begin{array}{l}
n=n_r+l +1,    \\
\tilde E +\omega ^2 = -\frac{\omega ^4r_g^2 }{16n^2 }=-\frac
{1}{a_\omega ^2 n^2}
\end{array}
\end{equation}
where we have the definition of the ``Bohr radius''
\begin{equation}
a_\omega =\frac {4}{\omega ^2r_g}.
\end{equation}
Note that real Bohr radius depends on Planck constant and does not
depend on any quantum number. However, the ``Bohr radius'' defined
here depends on the temporal quantum number $\omega $ and the
gravitational strength $r_g$. The quantization number $\breve h$ is
not involved. In the mean time, the WhittackerM function reduces to
the radial wave function of Hydrogen atom,
\begin{equation}
\begin{array}{l}
{\rm WhittackerM}(\kappa, \mu, z)= \\
e^{-\frac{r}{na_\omega } }r^{l+1} \mbox{ }_1F_1(-n+l+1, 2l+2,
\frac{r}{a_\omega }) \\
\propto r R_{nl}(r ).
\end{array}
\end{equation}

{\bf (ii) Fitting the planetary distances. } By using the wave
function of Hydrogen atom, Nottale, Schumacher, and Gay \cite{jin6}
gave an excellent fitting of the distributions of planetary
distances and planetary masses. The distances of the inner planets
(Mercury, Venus, the Earth, Mars, and the main mass peaks of the
asteroids belt) can be fitted by the formula
\begin{equation}
a_{nl}=\left (\frac{3}{2}n^2-\frac{1}{2}l(l+1)\right )a_\omega
\end{equation}
where $l=n-1$. The probability density peaks of Hydrogen atom wave
functions do identify with  the formula if we choose
\begin{equation}
\begin{array}{l}
\omega _{1}=6\times 10^{-7}\,{\rm m}^{-1}      \\
a_{\omega _{1}} =6.3 \times 10^{9}\,{\rm m}
\end{array}
\end{equation}
As revealed in Nottale, Schumacher, and Gay \cite{jin6}, the
distances of the outer planets (Jupiter, Saturn, Uranus, Neptune,
Pluto) and the one of the center mass of the inner planetary system
can be fitted by the formula too if we choose
\begin{equation}
\begin{array}{l}
\omega _{2}= \sqrt{5} \omega _{1}      \\
a_{\omega _{2}} = a_{\omega _{1}} /5
\end{array}
\end{equation}
This constitutes the hierarchical explanation of the quantized
planetary distances. Because the coefficients of our solvable
Lagrangian (14) do not involve time (static description), the
temporal operator has continual quantum number $\omega $ (see
(30)). However, the formulas (42) and (43) indicate that
$\omega $ is quantized too. The hierarchical structure of solar
system is consistently explained by the quantized values of $\omega
$ of the temporal operator. In the solar quantization theory of
Nottale, Schumacher, and Gay \cite{jin6}, however, not only the wave
function but also the ``Bohr radius'' involve a so-called universal
quantization constant $w_0$ and the hierarchical structure has to be
an additional assumption. Nottale, Schumacher, and Gay \cite{jin6}
also gave an excellent fragmentation explanation of solar planetary
masses based on the same wave function. }

\newpage
\pagenumbering{arabic}

\begin{center}
\large{
Prediction on Gravity Probe B
}
\end{center}

{\bf a. Introduction   }\\
\\
All direct tests of general relativity so far are performed on weak gravitational fields. Their theoretical calculations are all based on a set of coordinate system which has direct meaning of distance, angle, or time. This is possible only when spacetime is flat. Curved spacetime has no such coordinate system. Therefore, all classical tests of GR do not prove curved spacetime and the assumption of curved spacetime is still open for testification. However, the assumption must be false if GR fails to at least one classical test of weak gravity. Gravity probe B experiment (GP-B) provides such possibility of falsifying both Einstein field equation of GR and the assumption of curved-spacetime.

Relativists imagine that the whole universe plus its evolution is a curved 4-dimensional marble in, if possible, 5-dimensional flat space, similar to the 2-dimensional surface of a mountain in our daily 3-dimensional space. However, we human-being can not see the 4-dimensional marble in 5-dimensional space directly. In fact, the most accurate measuring equipments used to measure spatial distances and temporal intervals by scientists, are essentially electromagnetic wave, which is a  physical process in nature. Special relativity tells us that the wave lengths and frequencies vary with the reference frames by which we stand, and general relativity tells us that one physical process is affected by other. Therefore, Julian Barber proposed a revolutionary concept many decades ago that there is no time. Time is the impression of changes.

Einstein$^,$s general relativity and his agreement of curved spacetime assumption with mathematicians are solely based on his false imagination that two test masses of different initial speeds share the same local universal acceleration, which is called Einstein Equivalence Principle.
However, Galileo testified that two test masses share the same local universal acceleration only when they have identical initial speeds (in fact, zero initial speeds in his case). Now return to curved spacetime assumption. If there were one set of coordinate system $(t, x, y, z)$ (rectangular coordinates) on curved spacetime then the geodesic motion in terms of the coordinates is
$$
\frac{d^2x^i}{ds^2} = -\Gamma ^i_{jk} \frac{dx^j}{ds} \frac{dx^k}{ds} \,\,\,\,\,\,\, (1)
$$
which is simply:
$$
\mbox{ accelaration = connection times velocity times velocity }
\,\,\,\,\,\,\, (2)
$$
which says that local accelerations depend on test particle velocities.
Even though relativists can find their own definitions of velocities and accelerations on curved spacetime, the above geodesic motion must still hold and two masses of different initial speeds must not share the same local universal acceleration. Therefore, Einstein {\it universal acceleration imagination is contradictory to the assumption of curved spacetime!}
{\bf Real truth is that local accelerations are independent of the masses of test particles when they have identical initial speeds. This should be the so-called Equivalence Principle which is testified accurately. }  {\it Because of this principle, we can always use Einstein metric form to study gravity, which does not involve test particle mass. However, this does not mean that the metric has geometric meaning and describes curved spacetime.}

To resolve the difficulties of Big Bang theory, I proposed a model of the universe which is based on flat spacetime (He, 2006b). There is no cosmological expansion and the cosmological redshift is mainly gravitational one.
If spacetime is flat then we have inertial frames. He (2006b) proved the existence of the absolute and unique inertial frame of the universe. All other inertial frames are approximate ones and are freely falling (e.\,g., galactic clusters, individual galaxies, stars, sun, and earth), and form a hierarchical structures. The metric form of general relativity does not describe spacetime and is called refraction metric which like refraction medium is effective to curve light rays and massive bodies. All successful tests of GR are still true with respect to the refractive metric.

In summary, I present the following principles of flat spacetime gravity:
(1) Space and time are perceived and measured only by means of physical processes which are static to reference frames; (2) Freely falling mass (frame) in real surroundings always presents gravity and always has the background flat spacetime which were perceived and measured by the mass frame itself if it had approximately zero mass; (3) Spacetime is not curved and the metric form of general relativity (GR) is called refraction metric on flat spacetime which is effective to curve light rays. Similarly we have effective curvature, effective parallel displacement, etc. (4) The refraction metric (gravitational field) of a mass is generally obtained by holonomic or nonholonomic coordinate transformations.

This paper presents the other important principle. Einstein relativity is not complete. Mass and rotation are the basic components of nature and I propose the principle of universal gravitation due to rotation. Any part of a rotational mass generates an anisotropic force which is proportional to the product of the mass point and its squared relative velocity. Similar to Newton law, the force is proportional to the point mass on which the force is exerted and inversely proportional to the square of the distance between the two masse points. The new gravity is justified by the experimental detection of gravitomagnetic London moment by Tajmar {\it et al} (2006).
This principle and the above-mentioned principles are called Absolute Relativity.

These principles are to be testified by GP-B and other experiments. Fortunately, preliminary calculations based on the principles match observational facts consistently with solar system, galaxies, and the universe on the whole (He, 2005a, b, c, 2006a, b). Now we focus on Gravity Probe B experiment.
GR formulas of gyroscope precession in weak gravity are heuristically derived by applying a series of coordinate transformations.
This paper shows that the application of rotational coordinate transformation which corresponds to the addition of the above rotational gravity, is needed. First public peek at GP-B results shows that both geodetic and frame-dragging effects are larger than GR predictions by the amount of about 25 mas/yr (see Figure 1 which is taken from Everitt (2007)).
It is suggested that the gap can be covered by applying the above coordinate transformation, i.\,e., adding the rotational gravity. We wait for the final release of GP-B data analysis in the coming December and see if my calculation is confirmed.
Here I present my calculation of gyroscope precession in the following Section b. Section c is a simple discussion of motional gravity. Section d is conclusion.
\\
\\
{\bf b. Absolute Relativistic Formulation of Gyroscope Precession in Weak Gravity.}\\
\\
Adler and Silbergleit (2006) gave heuristic derivation of the formulas of gyroscope precession in weak gravity by applying a series of coordinate transformations. One of my new flat spacetime principles is that the refraction metric (gravitational field) of a mass is generally obtained by holonomic or nonholonomic coordinate transformations. This paper shows that the application of rotational coordinate transformation is needed. Here is the detailed calculation of gyroscope precession in weak gravitational field of earth.

My other principle says that the freely falling earth in real surroundings always presents gravity and always has the background flat spacetime which were perceived and measured by the earth itself if it had approximately zero mass and had no rotation. If earth had zero mass and no rotation then earth were the inertial reference frame which measured negligible gravity and we had the familiar Minkowski metric form,
$$
ds^2=c^2dT^2-d \stackrel{\rightarrow}{R}^2  \,\,\,\,\,\,\, (3)
$$
where $c$ is light speed and $(T, \stackrel{\rightarrow}{R}) $ is the Cartesian rectangular coordinate system which has direct meaning of distance and time. In reality, earth has mass and gravitation, and light rays no longer go straight lines in earth neighborhood.
The observationally verified Schwarzschild solution in isotropic coordinates (Adler and Silbergleit (2006), page 5) which described the refraction metric of earth if earth were stationary (no rotation), is, to first-order approximation
$$
\begin{array}{c}
ds^2 =\left(1-\frac{2GM}{c^2r}\right)c^2dt^2-
     \left(1+\frac{2GM}{c^2r}\right)d \stackrel{\rightarrow}{r}^2 \\
    \mbox{(stationary point mass)} \\
(4)
\end{array}
$$
where $G$ and $M$ are gravitational constant and earth mass, respectively, and $(t, \stackrel{\rightarrow}{r}) $ is exactly the same Cartesian rectangular coordinate system $(T, \stackrel{\rightarrow}{R}) $ according to my flat spacetime theory of gravity. That is, $(t, \stackrel{\rightarrow}{r}) $ ($\equiv (T, \stackrel{\rightarrow}{R}) $) were perceived and measured by means of physical processes which were static to the earth frame if earth had zero mass and no rotation. Therefore, $ds^2$ is no longer the physical proper distance on curved spacetime. All formal calculation of general relativity can be carried over to my theory, however, no geometric interpretation is allowed here. The coordinates $t, x, \phi $ are our physical time, physical distance, and physical angle respectively. However, they can not be measured directly because any measuring equipment (physical process) is also affected by the gravitation of the non-zero mass itself. Therefore,  my theory always describes the gravity of freely-falling mass based on the flat spacetime coordinate system which were perceived and measured by the mass itself if it had zero mass and no rotation.
According to He (2005c), the above Schwarzschild solution (4) is obtained from (3) by taking a nonholonomic coordinate transformation,
$$
d T = \surd{(1-\frac{2GM}{c^2r})}dt, \;\;
d R= \surd{(1+\frac{2GM}{c^2r})} dr   \,\,\,\,\,\,\, (5)
$$
By now we see no rotational gravity.

To proceed, however, Adler and Silbergleit (2006, page 5) missed the consideration of rotational coordinate transformation.
Earth has rotation with respect to the above inertial reference frame and it is the rotation that gives both geodetic and frame-dragging effects. Now we want to know the refraction metric of rotational earth described by people resting on the earth itself.
Based on the rotational frame, the earth has no rotation and we observe no rotational gravity. The coefficients of Schwarzschild solution describe Newton universal gravitation due to static (motionless) point mass. Therefore, the diagonal coefficients of the refraction metric observed by people resting on earth are the same as the ones of Schwarzschild solution. Scientists study earth gravity accurately and have never discovered my rotational gravity before. This is simply because their equipments rest on earth and observe earth gravity based on the earth rotational frame. GP-B experiment took place in outer space inside a satellite and its data analysis is based on the earth inertial frame or even the solar inertial frame. First public peek at Gravity Probe B results (GP-B) shows that both geodetic and frame-dragging effects are larger than GR predictions by the gap of about 25 mas/yr. This gap does result from rotational gravity as explained in the following.

The rotational frame, however, is not inertial one and the corresponding refraction metric must be different from Schwarzschild solution which describes gravity in inertial frame.
Therefore, the only way to change Schwarzschild solution (4) into the one which describes gravitational field with respect to rotational reference frame is to add off-diagonal components. I add the following off-diagonal component,
$$
f\frac{GM }{c^2r} \frac{s \omega }{c} \,\,\,\,\,\,\, (6)
$$
where $\omega $ is the angular speed of the rotating earth, $f $ is a constant (we will find out that $f\geq 4$ covers the gap from GR frame-dragging prediction),  and $(t_r, s, \phi _r , z)$ is the cylindrical coordinate system for the rotating earth while $(t, s, \phi  , z)$ is the corresponding cylindrical coordinate system for non-rotating earth: the inertial frame.
The off-diagonal term depends on both earth mass and rotational rate.
The refraction metric in the rotating frame is,
$$
\begin{array}{c}
ds^2 =\left(1- \frac{2GM}{c^2r_r }\right)c^2dt_r^2-
     \left(1+\frac{2GM}{c^2r_r}\right)(d s^2 +s^2d\phi _r^2+dz^2) +
f\frac{GM }{c^2r} \frac{s \omega }{c}cdt_r\,sd\phi
   \\
    \mbox{(in rotating earth frame)}   \\
(7)
\end{array}
$$
By now we still see no rotational gravity because the metric form (7) is based on the earth rotational frame and both observers and their equipments rest on earth surface. That is, there is no rotational gravity when people rest on earth and observe no earth rotation.

Note that the above formula when we choose $f=8$, is exactly the standard GR metric form which is used to describe gyroscopic procession with respect to earth inertial frame, i.\,e., the earth frame if it had no rotation.
First public peek at Gravity Probe B results (GP-B) shows that both geodetic and frame-dragging effects are larger than GR predictions by the gap of about 25 mas/yr. Therefore, standard GR metric form fails. It misses a rotational coordinate transformation as explained in the following.

Now we should return to the inertial frame, i.\,e., the earth frame if it had no rotation, and find its corresponding refraction metric which makes prediction on GP-B data.
Unfortunately, a century after Einstein proposed his special theory of relativity, we have not had a sound theory of relativistic rotation (Klauber, 2006)! Instead of taking a global rotational coordinate transformation, Adler and Silbergleit (2006, page 5) considered local Lorentz transformation. The transformation, however, applies to straight motion at constant speed only.
Because Lorentz transformation is homogeneous and describes homogeneous motion, it is highly suggestible to generalize Lorentz transformation into inhomogeneous one which describes global rotation.
Therefore, I have the following global rotational Lorentz transformation, to first order approximation in cylindrical coordinates,
$$
\begin{array}{l}
ct_r=\left(1+\frac{ 1 }{2} \left( \frac{s \omega }{c}\right)^2- H \frac{ M }{c^2r} \left( \frac{ s\omega }{c}\right)^2\right)
 \left(ct -\frac{s \omega }{c} s \phi  \right)    \\
s = s      \\
\phi _r=\left(1+\frac{ 1}{2} \left(\frac{s \omega }{c}\right)^2+ H \frac{ M }{c^2r} \left( \frac{ s\omega }{c}\right)^2\right)
 \left(\phi -\frac{\omega }{c} ct  \right)          \\
z = z  \\
(8)
\end{array}
$$
where $\omega $ and $M$ are the rotational rate and mass of earth, respectively, and $H$ is another constant associated with rotational gravity which plays the role of Newton gravitational constant $G$.
After the global transformation, the refraction metric (7) in rotational frame becomes finally the required metric in inertial frame,
$$
\begin{array}{ll}
ds^2 =&\left(1- \frac{2GM}{c^2r } - \frac{2HM}{c^2 r}\frac{s^2 \omega ^2}{c^2}\right)c^2dt^2
 -\left(1+\frac{2GM}{c^2r}+ \frac{2HM}{c^2 r}\frac{s^2 \omega ^2}{c^2} \right)(s^2d\phi ^2) \\
& -\left(1+\frac{2GM}{c^2r} \right)(d s^2 +dz^2)  \\
& +\left(  \frac{8GM }{c^2r} \frac{s \omega }{c} +f\frac{GM }{c^2r} \frac{s \omega }{c} \right) cdt\,sd\phi \\
  &  \mbox{(in inertial frame)} \\
& (9)
\end{array}
$$
Now we have the rotational terms, $\frac{HM}{c^2 r}\frac{s^2 \omega ^2}{c^2} $, which produce rotational gravity.
Therefore, negligible but rotating mass will still generate measurable gravitomagnetic effect. Martin Tajmar {\it et al} (2006) did observe such effect. Therefore, {\it our choice of rotational gravity is justifiable.}
GR calculation, however, is based on Einstein field equation whose resulting metric components have no rotational gravity.
GR fails to the explanation of the phenomena observed by Tajmar {\it et al} (2006). We know that diagonal components contribute to geodetic effect while
off-diagonal component contributes to frame-dragging effect.
The above rotational gravity will generate additional geodetic effect to cover the gap of GP-B data from GR prediction.

Now, we consider the off-diagonal component. The choice of $f\geq 4$ will generate additional frame-dragging effect to cover the gap of GP-B data from GR prediction. The global rotational Lorentz transformation has unsymmetrical denominators.

The above result applies to mass of small size only. Real mass (the earth) is a finite distribution of masses. We use superposition law to achieve its refraction metric.
Now we see my principle of universal gravitation of motions. Any part of a  rotational mass generates an anisotropic force which is proportional to the product of the part of mass and its squared relative velocity. Similar to Newton law, the force is proportional to the point mass on which the force is exerted and inversely proportional to the square of the distance between the masses:
$$
\mbox{universal gravity due to rotational mass} = H
\frac{M (v/c)^2 }{r^2}  \,\,\,\,\,\,\, (10)
$$
where $H$ is rotational gravitational-constant. The new gravity is justified by the experimental detection of gravitomagnetic London moment by Tajmar {\it et al} (2006).

Now we use superposition law and finally obtain the refraction metric for our earth,
$$
ds^2 =(1+\Phi )c^2dt^2-(1-\Phi )s^2d\phi ^2
 -(1-\Phi _1)(d s^2 +dz^2)  + h \, cdt\,sd\phi \,\,\,\,\,\,\, (11)
$$
where
$$
\Phi =\Phi _1 +\Phi _2, \,\,\, h =h _1 +h _2 \,\,\,\,\,\,\, (12)
$$
and
$$
\begin{array}{l}
\Phi _2 (\stackrel{\rightarrow}{r}) = - (2H/c^4)\int \rho (\stackrel{\rightarrow}{r}^\prime )  v ^2 (\stackrel{\rightarrow}{r}^\prime )  d^3 \stackrel{\rightarrow}{r}^\prime /|\stackrel{\rightarrow}{r}-\stackrel{\rightarrow}{r}^\prime |   ,  \\
h_2 (\stackrel{\rightarrow}{r}) =(fG/c^3) \int
\rho (\stackrel{\rightarrow}{r}^\prime )  v (\stackrel{\rightarrow}{r}^\prime )    d^3 \stackrel{\rightarrow}{r}^\prime /|\stackrel{\rightarrow}{r}-\stackrel{\rightarrow}{r}^\prime |  ,
  \\
\Phi _1(\stackrel{\rightarrow}{r}) =-(2G/c^2) \int
\rho (\stackrel{\rightarrow}{r}^\prime )d^3 \stackrel{\rightarrow}{r}^\prime /|\stackrel{\rightarrow}{r}-\stackrel{\rightarrow}{r}^\prime |  ,   \\
h_1 (\stackrel{\rightarrow}{r}) =(8G/c^3) \int
\rho (\stackrel{\rightarrow}{r}^\prime )  v (\stackrel{\rightarrow}{r}^\prime )    d^3 \stackrel{\rightarrow}{r}^\prime /|\stackrel{\rightarrow}{r}-\stackrel{\rightarrow}{r}^\prime |  \\
(13)
\end{array}
$$
where $\rho $ is earth mass density and $ v $ is its  linear velocity  due to earth rotation.
The terms with subscript $_1$ are exactly the solution of the linearized Einstein field equation and the ones with subscript $_2$ are my suggestion. When applied to the rotational earth and the orbiting gyroscopes of GPB, GR predicts smaller results of both geodetic and frame-dragging effects by the amount of about 25 mas/yr. This gap can be simply covered by adding the rotational gravity $\Phi _2 $ and the off-diagonal term in non-inertial rotational frame
$h _2$. This suggestion is left for further investigation.
\\
\\
{\bf c. Is Rotational Gravity a Consistent Assumption? }\\
\\
Since we have gravitational field (11) (refraction metric form), we can study the motion of a test particle or a gyroscope in the field.  The metric form tells us that the gravity due to mass, $\Phi _1$, is isotropic while the gravity due to rotation, $\Phi _2$, is anisotropic. Because rotational gravity is larger in equator direction, we had a simple explanation to why solar system is planar and all rotations are approximately in the same plane. Comparing the formula (10) with Newton law, we know that the amplitude of rotational gravity is proportional to the squared relative velocity while the amplitude of Newton gravity is proportional to the  gravitational constant:
$$
H(v/c) ^2 \, \propto \, G  \,\,\,\,\,\,\, (14)
$$

1. For GP-B experiment, the difference is
$$
\sim H \times 10^{-13} \,\,\, \propto \, \,\, \sim 10^{-11}  \,\,\,\,\,\,\, (15)
$$
which is comparable to the first result of GP-B data: 25 mas/yr $\propto $ 6606 mas/yr, if $H \approx 1$. To get the real value of geodetic effect, we need calculate parallel transport of gyroscope. The calculation, however, is very complicated. Here, I look for simple method to achieve approximate result. Firstly, note that the rotational gravity $\Phi _2$ is not isotropic (see formula (11)). Accordingly, the Christoffel symbols between (15) and (16) of page 6 in Adler and Silbergleit (2006), i.,e., AS (2006) in the following, would be $\Phi$ sometimes and $\Phi _1$ in other times.   However, the terms with factor $\alpha $ must always be $\Phi$. Therefore, the instantaneous value of geodetic precession $\Omega _G$ in (18) of AS (2006) must be,
$$
\stackrel{\rightarrow}{\Omega _G} \equiv \frac{\alpha }{2} \nabla \Phi \times
\stackrel{\rightarrow}{V} + \frac{2\gamma }{2} \nabla \Phi _{?} \times
\stackrel{\rightarrow}{V}
\,\,\,\,\,\,\, (16)
$$
For the special configuration of GP-B experiment, however, $\Phi _{?} $ must be $\Phi _{1}$. Therefore, if we find the amplitude of rotational gravity (compared to newton gravity), only one third of the amplitude contributes to our geodetic effect ($\alpha = \gamma =1$). Now I calculate the amplitude of rotational gravity. The rotational gravity (10) of earth to a unit mass at distance $r$ from earth center and in earth polar direction is approximately,
$$
\frac{H M }{r^2} \left(\frac{R\omega }{c}\right)^2 \frac{2 }{5} \,\,\,\,\,\,\, (17)
$$
where $R$ is earth radius.
The factor 2/5 corresponds to the amplitude of rotational gravity in earth polar direction. It is very hard to calculate the gravity in other radial direction. We expect an averaging factor of 2/4.5 for all directions. Now we are ready to approximate geodetic contribution due to rotational gravity.
$$
H\left(\frac{R\omega }{c}\right)^2 \frac{2 }{4.5} =H \times 0.106 \times 10^{-11}  \,\, {\rm N\, m}^2/{\rm kg}^2   \,\,\,\,\,\,\, (18)
$$
Therefore, rotational gravity for earth is less than Newton gravity by $0.106/6.67 = 0.016$, if we assume $H \approx 1$. The detailed analysis (formula (16)) indicates that only one third of rotational gravity contributes to GP-B experiment. Finally, the contribution (i.e., ur prediction on geodetic effect is larger than GR by the following amount) is
$$
H\times \frac{0.016}{3} \times 6606 = 35 \times H \,\,\mbox{mas/yr}   \,\,\,\,\,\,\, (19)
$$
Symmetric consideration suggests that the factor $f$ in (9) and (13) is 4. Therefore, our prediction on frame-dragging effect is larger than GR by about 19.5\,mas/yr.

2. For sun,
$$
\sim H\times 10^{-13} \,\,\, \propto \,\,\, \sim
     10^{-11}  \,\,\,\,\,\,\, (20)
$$
Therefore, the usually quoted value of mass of sun by scientist is not its real mass.  It is the real mass plus its `rotational mass$^,$ due to rotation. The mass of sun is less than the quoted value $5.98 \times 10^{30}$ by at most one percent, if we assume $H \approx 1$.

3. We know that rotational gravity depends on velocity. It is interesting that spiral galaxies have masses towards their centers exponentially but the velocity of their components is approximately constant. Therefore, galaxy constant rotational velocity is mainly due to their rotational gravity!

I do not seek further evidences. We spend billions of dollars to look for dark matter and dark energy. Why not spend millions of dollars to repeat the experimental detection of gravitomagnetic London moment performed by Tajmar {\it et al} (2006)?
\\
\\
{\bf d. Conclusion   }\\
\\
Standard calculation of gyroscope precession is based on the curved spacetime assumption and Einstein field equation. The assumption encounters tremendous difficulty in the explanation of galaxies, galaxy clusters, and the universe on the whole while classical tests of GR do not prove curved spacetime. In curved spacetime no coordinate system has direct meaning of length, time, and angle.
However, the curved spacetime assumption must be false if GR fails to at least one classical test of weak gravity. Gravity probe B experiment (GP-B) provides such possibility. First public peek at GP-B results shows that both geodetic and frame-dragging effects are larger than GR predictions by the amount of about 25 mas/yr (see Figure 1).
It is suggested in this paper that the gap can be covered by applying rotational  coordinate transformation which corresponds to introducing rotational gravity. We wait for the final release of GP-B data analysis in the coming December and see if my calculation is confirmed.
\\
\\
\large {{\bf Figure caption } } \normalsize \,\,\,\,
Figure 1. Glimpses of frame-dragging effect taken from {\it Gravity Probe B: Interim Report and First Results }, a talk given by Francis Everitt at APS meeting
2007 in Jacksonville, Florida. See the website {\it www.einstein.stanford.edu.}
\\
\\
\\
Everitt, F.: 2007, {\it Gravity Probe B: Interim Report and First Results } in APS meeting, www.einstein.stanford.edu
\\ He, J.: 2005a, astro-ph/0510535 %
\\ He, J.: 2005b, astro-ph/0510536 %
\\ He, J.: 2005c, astro-ph/0512614v3 %
\\ He, J.: 2006a, astro-ph/0604084 %
\\ He, J.: 2006b, astro-ph/0605213 %
\\ Klauber, R.\,D.: 2006, gr-qc/0604118 %
\\ Tajmar, M., Plesescu, F., Marhold, K., \& de Matos, C.\,J.: 2006,
gr-qc/0603033 %

\end{document}